\documentclass[pre,preprint,preprintnumbers,amsmath,amssymb]{revtex4}
\usepackage{graphics,epsfig,epstopdf,stmaryrd}
\begin {document}
\title{Self-consistent theory for the linear and nonlinear  propagation of a sinusoidal electron plasma wave. Application to stimulated Raman scattering in a non-uniform and non-stationary plasma}
\author{Didier B\'enisti}
\affiliation{ CEA, DAM, DIF F-91297 Arpajon, France.}
\email{didier.benisti@cea.fr} 
\date{\today}
\begin{abstract}
In this paper, we address the theoretical resolution of the Vlasov-Gauss system from the linear regime to the strongly nonlinear one, when significant trapping has occurred. The electric field is that of a sinusoidal electron plasma wave (EPW) which is assumed to grow from the noise level, and to keep growing at least up to the amplitude when linear theory in no longer valid (while the wave evolution in the nonlinear regime may be arbitrary). The ions are considered as a neutralizing fluid, while the electron response to the wave is derived by matching two different techniques. We make use of a perturbation analysis similar to that introduced to prove the Kolmogorov-Arnold-Moser theorem, up to amplitudes large enough for neo-adiabatic results to be valid. Our theory is applied to the growth and saturation of the beam-plasma instability, and to the three-dimensional propagation of a driven EPW in a non-uniform and non-stationary plasma. For the latter example, we lay a special emphasis on nonlinear collisionless dissipation. We provide an explicit theoretical expression for the nonlinear Landau-like damping rate which, in some instances, is amenable to a simple analytic formula. We also insist on the irreversible evolution of the electron distribution function, which is nonlocal in the wave amplitude and phase velocity. This makes trapping an effective means of dissipation for the electrostatic energy, and also makes the wave dispersion relation nonlocal. Our theory is generalized to allow for stimulated Raman scattering, which we address up  to saturation by accounting for plasma inhomogeneity and non-stationarity, nonlinear kinetic effects, and interspeckle coupling. 
\end{abstract}
\maketitle
\section{Introduction}
Solving the Vlasov-Gauss system is one of the key issues in plasma physics. Recently, it was addressed mathematically by C. Mouhot and C. Villani~\cite{villani} who proved that, provided that its initial value was small enough, the electrostatic potential decreased down to zero (or, more precisely, they proved that the charge density converged towards its space averaging). This was quite a mathematical achievement, since C. Villani was awarded the 2010 Fields Medal ``for his proofs of nonlinear Landau damping and convergence to equilibrium for the Boltzmann equation''. Although their work was dubbed nonlinear Landau damping, because it solved  Vlasov equation without resorting to linearization, it only applied to small amplitude potentials, i.e., to what a physicist would call the linear regime. Moreover, the mathematical theory of Ref.~\cite{villani} did not provide the damping rate and could not address the Coulomb potential. This shows how difficult it is to obtain rigorous results as regards the nonlinear resolution of the Vlasov-Gauss system. \\

In this paper, we do not try to build a mathematical theory, but we derive accurate results which do hold in a regime where linear theory is no longer valid and, hopefully, which are new and of practical interest. More precisely, this article is mainly devoted to the three following points :
\begin{enumerate}
\item [(i)] In several previous publications, we obtained theoretical results pertaining to the nonlinear motion of electrons acted upon by a space and time varying potential, associated with an electrostatic wave. Here, we want to give a synthetic and quick overview of these results, in order to put forward our analytic derivation of the charge density induced by the potential.  Our method rests on a very precise description of the nonlinear electron motion, obtained by connecting two different perturbative techniques which are detailed below, and which are valid for two opposite ranges (small or large) of amplitudes. While the large amplitude results are derived whether the wave grows or decays, the small amplitude analysis only holds when the wave amplitude (as seen by the electrons) increases. The latter restriction, that cannot be alleviated, corresponds to the most general situation since a wave usually grows before entering a strongly nonlinear regime. We also restrict here to sinusoidal waves, although the application of our method to more general potentials poses no conceptual difficulties (but may not be straightforward on a technical point of view). Moreover, we need to recall our previous theoretical results, in order to use them for the two following applications, which constitute the two new results of this paper. 

\item[(ii)] We provide, for the first time, a purely theoretical resolution of the nonlinear growth and saturation of the cold beam-plasma instability, which is certainly the oldest and most simple problem of nonlinear plasma kinetic theory. This allows to derive  the level and time of saturation of this instability more rapidly than by resorting to particle simulations. This is needed when addressing three-dimensional problems of large extent, or for specific technical applications as those related to a traveling wave tube~\cite{top}. 

\item[(iii)] We derive a set of coupled differential equations which, for the first time, describe stimulated Raman scattering~(SRS) up to saturation, in a non-uniform and non-stationary plasma, by accounting for nonlinear kinetic effects and interspeckle coupling. SRS is still an issue for laser fusion, and providing a model able to predict Raman reflectivity in a fusion plasma was the main motivation for the work presented in this article. Consequently,  our hypotheses will often be justified by their relevance to SRS.
\end{enumerate}
Note that points (ii) and (iii) are strongly  correlated. Indeed, SRS efficiency depends a lot on the value of the EPW Landau damping rate, which nonlinearly decreases due to electron trapping. The very same trapping mechanism saturates the beam-plasma instability. \\

In the whole paper, we assume that the ions may be considered as a neutralizing fluid, which is true when SRS is dominated by kinetic effects~\cite{rousseaux06}. As regards the electron motion, it is derived from a direct resolution of Newton equation.  As stated above, for the sake of simplicity we restrict to sinusoidal waves, which  is usually   a good approximation for SRS. However, the technique we use to solve the electron dynamics applies to any potential. We also restrict to the situation when the wave grows from noise and keeps growing until it enters the nonlinear regime (while, afterwards, its evolution may be arbitrary). As already indicated above, this hypothesis cannot be avoided, but it corresponds to the most common situation, usually a wave grows before entering the nonlinear regime. \\

We first address the electron response to the wave when the EPW and the plasma are uniform. For small amplitudes, we make use of a perturbation analysis based on canonical transforms~\cite{goldstein,lieberman}, which is similar to that introduced to prove the Kolmogorov-Arnold-Moser (KAM) theorem ~\cite{KAM}. Such a perturbative scheme happens to be useful to accurately solve many problems of plasma physics (see for example Refs.~\cite{brb1,brb2,brb3,brb4}), even way beyond the range in physics parameters that would correspond to the KAM theorem itself. Indeed, in Ref.~\cite{locality}, a perturbative expansion was used to derive the statistical properties of a chaotic dynamics and to prove that this dynamics could be modeled by a diffusion equation~\cite{benisti98}. Similarly, in this paper we show that, for a growing wave, perturbative results are valid up to unexpectedly large amplitudes, when the electrons may be considered as trapped (see Section~\ref{IIA1}). For a wave with increasing amplitude, the small parameter of the perturbative expansion is of the order of $\omega_B/\gamma$, where $\gamma$ is the wave growth rate and where $\omega_B$ is the so-called bounce frequency, $\omega_B \equiv  \sqrt{eE_0m/k}$, $E_0$ being the electric field amplitude, $k$ the wavenumber, and  $-e$ and $m$ are, respectively, the electron charge and mass. Hence, a perturbative analysis is only valid for small enough values of $\omega_B/\gamma$. It yields an approximate constant for the electron motion (see Section~\ref{IIA1}), which allows to straightforwardly derive the electron distribution function in the corresponding variable.

In the opposite regime, when $\omega_B /\gamma\gg 1$, the dynamics changes at a rate much smaller than the typical period of an electron orbit. Then, the adiabatic or neo-adiabatic theories described in Refs~\cite{lenard,tim,cary,hanna,nei,ten,vas} apply, which allows to precisely derive the electron distribution function in the so-called action variable, defined by Eq.~(\ref{n1000}).

Now, in Refs.~\cite{benisti07,action} we showed the quite unexpected result that there existed a range of values in $\omega_B/\gamma$ where the neo-adiabatic and the KAM-like perturbation theories were both very accurate, provided that the perturbative expansion was led up to a high enough (11th) order. Using this property, we derive the electron response to the wave (and, in particular, the charge density induced by this wave) for two very different physical problems. 

First, we address the situation when  the initial distribution function is of Dirac type, $f_0(v)=\delta (v-v_0)$, and when $\gamma/k\vert v_\phi-v_0\vert \agt 1$, where $v_\phi\equiv \omega/k$ is the EPW phase velocity. The latter condition is fulfilled for the cold beam-plasma instability~\cite{oneil71}, so that inserting the analytical expression found for the charge density into Gauss law yields an algebraic equation for the nonlinear growth rate, $\gamma$, of this instability. Solving for $\gamma$, we derive in Section~\ref{IIA2} the nonlinear evolution of the instability until (and even beyond) saturation. 

Second, we consider a driven wave that grows in an initially Maxwellian plasma so slowly that,  $\gamma/kv_{th}Ê\alt 0.1$, $v_{th}$ being the electron thermal velocity. Under these hypotheses, we show the unexpected result that the electron response to the wave abruptly changes from perturbative to adiabatic when $\int \omega_B dt \agt5$. Inserting, again, the analytical expression found for the charge density into Gauss law, we derive in Section~\ref{IIB} a  first order differential equation for the time variations of the EPW amplitude, under the action of the drive. In particular,  we show that the abrupt change in the electron response leads to a phase-like transition in the EPW properties.  \\

The results derived for a homogeneous electron plasma wave propagating in a uniform plasma are generalized to allow for a three-dimensional (3-D) variation of the wave amplitude, frequency and wavenumber, and a non-uniform and non-stationary plasma. However, we assume that no random, nor periodic, density fluctuations has grown so that the electron plasma wave is not subjected to Anderson-like localization~\cite{escande} ; we still consider a propagating wave. 
By resorting to a variational formalism, we derive a first order envelope equation, that accounts for collisionless dissipation in a very accurate fashion, in the linear and nonlinear regimes. In particular, we provide an explicit and simple theoretical expression (and even an analytic formula) for the nonlinear Landau-like damping rate. To account for dissipation, the Lagrangian density used to describe the EPW propagation is necessarily nonlocal. In particular, we insist here on the fact that the adiabatic distribution function is not local, and that it evolves in an irreversible fashion due to separatrix crossing. Then, the EPW nonlinear dispersion relation is also nonlocal.  We explicitly derive  this dispersion relation in Section~\ref{IIIB} and discuss it in detail, especially as regards the nonlinear frequency shift, $\delta \omega$. \\

By letting a laser light and a backscattered electromagnetic wave propagate together with the EPW, we allow for stimulated Raman scattering. Theoretically, we derive a set of coupled differential equations that yield the 3-D space and time evolution of SRS in a non-uniform and non-stationary plasma. It accurately accounts for the so-called kinetic inflation i.e., the fact that SRS grows more rapidly in the nonlinear regime than in the linear one, due to the decrease of collisionless dissipation. Our equations are valid up to SRS saturation, which is caused by the frequency shift or by the unstable growth of secondary electrostatic modes. Indeed, $\delta \omega$ entails a phase mismatch between the EPW and the laser drive, and also leads to the EPW self-focusing, which was shown in Refs.~\cite{yin,srs3D} to stop the growth of SRS. Both effects are explicitly accounted for in our envelope equations. Moreover, a large amplitude plasma wave is known to be unstable against the trapped particles instability~\cite{kruer,dodin1}, which leads to the growth of secondary electrostatic modes. In Ref.~\cite{friou}, Raman scattering was shown to saturate when the amplitude of the fastest growing mode overtakes that of the SRS-driven plasma wave. In order to account for this saturation mechanism, we calculate the linear growth of the secondary modes together with the amplitude variations of the EPW, as derived from the resolution of our envelope equations. \\

This paper is organized as follows. In Section~\ref{II} we derive the electron response to a homogeneous sinusoidal electron plasma wave that grows in a uniform plasma.  From this derivation, we provide a theoretical description of the growth and saturation of the beam-plasma instability, and we discuss the linear and nonlinear properties of a driven EPW growing in an initially Maxwellian plasma. In Section~\ref{III}, by resorting to a variational formalism, we generalize our results regarding wave propagation to a 3-D geometry and to a non-uniform and non-stationary plasma. In Section~\ref{IV}, the envelope equation for the EPW is coupled with that of a laser light and of an electromagnetic backscattered wave in order to describe stimulated Raman scattering, up to saturation, by accounting for nonlinear kinetic effects together with the plasma density variations. Section~\ref{V} summarizes and concludes our work. 

\section{Homogeneous sinusoidal wave in a uniform and stationary plasma}
\label{II}
\subsection{Electron response to a rapidly varying wave and application to the beam-plasma instability}
\label{IIA}
\subsubsection{Electron response to a rapidly varying wave}
\label{IIA1}
In this Section, we only consider homogeneous sinusoidal electrostatic waves, whose electric field reads,
\begin{equation}
\label{eq1}
E \equiv E_0(t) \sin(\varphi).
\end{equation}
From the eikonal, $\varphi$, we define the wavenumber $k\equiv \partial_x \varphi$ and the wave frequency, $\omega \equiv -\partial_t \varphi$. If the electrostatic field is sinusoidal, so is the charge density, which reads,
\begin{equation}
\label{eq2}
\rho \equiv \rho_s\sin(\varphi)+\rho_c \cos(\varphi).
\end{equation}
From the very definition of the charge density, one straightforwardly finds that $\rho_c = -2ne\langle \cos(\varphi) \rangle$ and $\rho_s=-2ne\langle \sin(\varphi) \rangle$, where $\langle . \rangle$ stands for a local statistical averaging. For any function $g(\varphi)$,
\begin{equation}
\label{eq3}
\langle g \rangle \equiv (2\pi)^{-1} \int_{\varphi_0-\pi}^{\varphi_0+\pi} \int_{-\infty}^{+\infty} f(\varphi,v)g(\varphi)dvd\varphi,
\end{equation}
where $f$ is the electron distribution function, normalized to unity. Note that, when the wave is homogeneous, $\langle \cos(\varphi) \rangle$ and $\langle \sin(\varphi) \rangle$ are independent of $\varphi_0$. 

 In this Section, we focus on the derivation of $\langle \sin(\varphi) \rangle$, which yields the time variations of $E_0$. The derivation of $\langle \cos(\varphi) \rangle$ is postponed to Section~\ref{IIIB}~when we address the EPW dispersion relation. When the wave amplitude is small, we provide a perturbative expansion of $\langle \sin(\varphi) \rangle$ by making use of canonical transformations, similar to those introduced to prove the KAM theorem~\cite{KAM}. Canonical perturbation theory is a very well-known technique that can be found in textbooks~\cite{goldstein,lieberman}. Let us quickly explain here how it may be applied to the derivation of $\langle \sin(\varphi) \rangle$. The dynamics of electrons, acted upon by the electric field $E$ defined by Eq.~(\ref{eq1}), derives from the Hamiltonian,
 \begin{equation}
\label{n1}
H=k(v-v_\phi)^2/2-(eE_0/m)\cos(\varphi),
\end{equation}
for the canonically conjugated variables $\varphi$ and $v \equiv dx/dt$. It is well-known~\cite{goldstein} that one may find a new pair of canonically conjugated variable $(\varphi',v')$ by using a generative function, $F(\varphi,v',t)$, such that,
\begin{eqnarray}
\label{n2}
\varphi'&=&\varphi+\frac{\partial F(\varphi,v')}{\partial v'}, \\
\label{n3}
v&=&v'+\frac{\partial F(\varphi,v')}{\partial \varphi} ,
\end{eqnarray}
and, in these new variables, the dynamics derives from the Hamiltonian $H' = H+\partial_tF$. Now, if one is able to find $F$ such that $H'$ only depends on $v'$, the electron motion is readily solved. Indeed,
\begin{eqnarray}
\label{n4}
v'(t)&=&v'(0) \\
\label{n5}
\varphi'(t)&=&\varphi'(0)+v'(0)t-\int_0^t\partial_{v'}H'(v',t')dt'.
\end{eqnarray}
Since $v'$ is conserved, its distribution function is just the initial one. Moreover, if we assume that, at $t=0$, the wave amplitude is vanishingly small, $v'(0)=v(0)$. Hence, the distribution in $v'$ is just the unperturbed velocity distribution function, which we denote by $f_0$. Similarly, if we assume that the initial distribution in $\varphi$ is uniform, we find from Eq~(\ref{n4}) that the distribution in $\varphi'$ is also uniform. Consequently, it straightforwardly follows from Eq.~(\ref{n2}) that,
\begin{equation}
\label{n6}
\langle \sin(\varphi) \rangle = \frac{1}{2\pi} \int_{-\infty}^{+\infty} f_0(v')\int_{-\pi}^{\pi}\sin[\varphi'-\partial_{v'}F(\varphi',v')]d\varphi'dv',
\end{equation}
which yields an explicit expression for $\langle \sin(\varphi) \rangle$ once $F$ is known. Now, because the dynamics defined by Hamiltonian $H$ is not integrable, it is not possible to find $F$ such that $H'=H+\partial_t F$ is exactly independent of $\varphi'$. However, it is possible to find an expansion for the generative function, $F\equiv \sum_{j=1}^N \varepsilon^jF_j$ (where $N$ is the order at which the expansion has been led, and $\varepsilon$ is the so-called small parameter of the perturbative analysis), such that the $\varphi'$-dependent term in $H'$ is of the order of $\varepsilon^{N+1}$. Then, Eqs.~(\ref{n4}) and (\ref{n5}) are only correct up to a term of the order of $\varepsilon^{N+1}$, and the same is true for Eq.~(\ref{n6}) which, when $\varepsilon \ll 1$,  yields a very precise expression of $\langle \sin(\varphi) \rangle$. Clearly, since the perturbative scheme aims at eliminating the potential part in Hamiltonian $H$, $\varepsilon$ should be proportional to the wave amplitude or, similarly, to $\omega_B^2 = eE_0k/m$. Moreover, if a wave grows very quickly, it may reach large amplitudes by the time the electrons could move significantly. Hence, for larger growth rates, $\gamma$, the electron motion remains little affected by the wave, and may be accurately approximated by a perturbative expansion, up to larger amplitudes.  Therefore, $\varepsilon$ should decrease with $\gamma$. Actually, since the electrons follow orbits that significantly depart from ballistic ones after a time of the order of $\omega_B^{-1}$, one expects $\varepsilon \sim (\omega_B/\gamma)^2$.

 The perturbative estimate of $\langle \sin(\varphi) \rangle$ has been derived in Ref.~\cite{benisti07} for a growing wave ($\gamma>0$), using a routine written with the maple\textregistered~software package. At lowest order in the variations of $\gamma$, $\omega$, and $k$, it reads
 \begin{equation}
\label{eq4}
\langle \sin(\varphi) \rangle \equiv \int_{-\infty}^{+\infty}Êf_0(v_0)S(v_0-v_\phi)dv_0,
\end{equation}
where the kernel $S$ is,
\begin{equation}
\label{eq5}
S(v_0-v_\phi)=\sum_{j=0}^{N} s_{2j+1}(v_0-v_\phi),
\end{equation}
for an expansion up to order $2N+1$. The expressions of the $s_n$'s may be found in Ref.~\cite{benisti07} up to $s_{11}$. Let us recall here that,
\begin{eqnarray}
\label{eq6}
s_1&=&\omega_B^2\frac{\gamma [k(v_0-v_\phi)]}{(\gamma^2+[k(v_0-v_\phi)]^2)^2},\\
s_3 &=& -\omega_B^6 \frac{3(43\gamma^4-26\gamma^2[k(v_0-v_\phi)]^2-5[k(v_0-v_\phi)]^4)[k(v_0-v_\phi)]\gamma}{8\left\{(9\gamma^2+[k(v_0-v_\phi)]^2)^2(\gamma^2+[k(v_0-v_\phi)]^2)^4\right\}}.
\end{eqnarray}
For values of $v_0$ such that $\vert v_0-v_\phi \vert \ll \gamma/k$, $\vert s_3/s_1\vert \propto (\omega_B/\gamma)^4$, and the same is true at any order $n$, $\vert s_{2n+1}/s_{2n-1}\vert \propto (\omega_B/\gamma)^4$ when  $\vert v_0-v_\phi \vert \ll \gamma/k$. We conclude that the perturbative expansion of $\langle \sin(\varphi) \rangle$ is only accurate for small enough values of $\omega_B/\gamma$, as expected from the previous discussion on the value for the small parameter, $\varepsilon$, of the perturbative expansion. 

Hence,  for larger growth rates the perturbative estimate of $\langle \sin(\varphi) \rangle$ is valid up to larger wave amplitudes.  Actually, perturbative results become particularly interesting when
\begin{equation}
\label{eq7}
\gamma/k\vert v_0-v_\phi \vert \agt 1,
\end{equation}
which we define as the condition for a rapidly varying wave. Indeed, when Eq.~(\ref{eq7}) is fulfilled, the numerical study of Ref.~\cite{action}~showed that $S(v_0-v_\phi)$ was very accurately estimated by its perturbative expansion up to amplitudes such that, for most electrons with initial velocity $v_0$,  
\begin{equation}
\label{eq9}
\zeta \equiv (kH+\omega_B^2)/2\omega_B^2
\end{equation}
was less than unity. From the definition Eq.~(\ref{n1}) of $H$, these electrons may be considered as trapped in the wave potential.

Now, it is well-known that the period of trapped orbits, $T$, away from the separatrix,  is close to $2\pi/\omega_B$ [which is easily recovered by replacing $\sin(x)$ by $x$ in $H$]. Moreover, it is also well-known that, when $T$ is much less than the typical time of variation of the Hamiltonian i.e., when $\gamma/\omega_B$ is small enough, adiabatic theory~\cite{lenard,goldstein, lieberman} guarantees the near conservation of the action which, for trapped electrons, is defined by,
\begin{equation}
\label{eq10}
I \equiv (4\pi)^{-1}Ê\oint vd\varphi,
\end{equation}
the integral being calculated over a so-called frozen orbit, i.e., for a fixed a value of $H$. For a sinusoidal wave, there exists an explicit analytical formula for $I$, 
\begin{equation}
\label{eq11}
I=\frac{4v_{tr}}{\pi}Ê\left[K_2(\zeta)+(\zeta-1)K_1(\zeta)Ê\right],
\end{equation}
where $v_{tr}Ê\equiv \omega_B/k$ and where $K_1$ and $K_2$ are, respectively, the elliptic integrals of first and second kind~\cite{abramowitz}. Once the action remains nearly conserved, it is more suited to study the electron motion in action-angle variables, where the angle, $\theta$, is canonically conjugated to $I$. Using the very definitions of $I$ and $\theta$~\cite{lieberman}, one straightforwardly finds, 
\begin{eqnarray}
\nonumber
\sin(\varphi)&=&2\sin(\varphi/2)\cos(\varphi/2) \\
\label{n101}
&=& 2 \sqrt{\zeta}~\text{sn}\left[Ê\left. \frac{2 K_1 \theta}{\pi}Ê\right \vert \zeta\right]\times \text{dn}\left[Ê\left. \frac{2 K_1 \theta}{\pi}\right\vert \zetaÊ\right],
\end{eqnarray}
where $Ê\text{sn}(u\vert \zeta)$ and $\text{dn}(u \vert \zeta)$ are Jacobian elliptic functions~\cite{abramowitz}. Using Eq.~(\ref{n101}), one may easily derive $\langle \sin(\varphi) \rangle$ for a set of trapped particles  which all have the same action, $I$, that may be considered as a constant. Now, as proved in Ref.~\cite{action}, the latter distribution in $I$ is indeed obtained for an initial Dirac distribution in velocity, $f_0(v)=\delta(v-v_0)$, provided that Eq.~(\ref{eq7}) is fulfilled, and once most electrons have been trapped in the wave potential (say for $E_0\geq E_M$). Moreover, unless $\gamma$ dramatically increases after trapping, $\gamma/\omega_B$ is small enough for the action to be nearly conserved whenever $E_0\geq E_M$. In order to be consistent with Eq.~(\ref{eq4}), let us denote by $S(v_0-v_\phi)$ the value of $\langle \sin(\varphi) \rangle$ corresponding to an initial Dirac distribution. Then, by averaging Eq.~(\ref{n101}) over $\theta$ one finds (see Ref.~\cite{action}), 
\begin{equation}
\label{eq14}
S(v_0-v_\phi)=S_M\frac{K_1^2(\zeta_M)}{K_1^2(\zeta)} \sqrt{\frac{q}{q_M}}\frac{1-q_M}{1-q} \cos\left[\int_{t_M}^t \frac{\pi\omega_B(t')}{2K_1[\zeta(t')]}dt' \right],
\end{equation}
where
\begin{equation}
\label{eq15}
q \equiv e^{-\pi K_1(1-\zeta)/K_1(\zeta)},
\end{equation}
and $q_MÊ\equiv q(\zeta_M)$, $\zeta_M$ being defined by Eq.~(\ref{eq11}) with $E_0=E_M$.

Hence, from Eq.~(\ref{eq14}) we know how to calculate $S(v_0-v_\phi)$ when $\gamma/\omega_B$ is small enough by making use of the adiabatic approximation. Moreover, when $\gamma/\omega_B$ is large enough, we have the perturbative estimate of $S(v_0-v_\phi)$ given by Eq.~(\ref{eq5}). Now, in Ref.~\cite{action} we proved the quite unexpected result that, when Eq.~(\ref{eq7}) was fulfilled, there existed a range in $\gamma/\omega_B$ when the adiabatic and the perturbative estimates of $S(v_0-v_\phi)$ were both accurate. More precisely, by leading the perturbative expansion Eq.~(\ref{eq5}) to order 11, one obtains an accurate estimate of $S(v_0-v_\phi)$ up to the amplitude, $E_M$, when it reaches its first maximum.  Moreover, when $E_0\geq E_M$, Eq.~(\ref{eq14}) provides an excellent approximation of $S(v_0-v_\phi)$. Therefore, we known how to derive an explicit expression for $S(v_0-v_\phi)$ whatever the wave amplitude. Using Eq.~(\ref{eq4}), this yields an accurate expression for $\langle \sin(\varphi) \rangle$ for any initial distribution $f_0(v_0)$ with a finite support such that, for any $v_0$ within that support, $\gamma/k(v_0-v_\phi)$ is larger than unity, at least when $E_0\leq E_M$.

We are now going to use this result in order to derive the first analytical description of the nonlinear growth and saturation of the cold beam-plasma instability. 

\subsubsection{Application to the cold beam-plasma instability}
\label{IIA2}
The nonlinear growth and saturation of the cold beam-plasma instability may be viewed as the oldest and most simple problem in nonlinear kinetic plasma theory. It corresponds to the initial distribution function,
\begin{equation}
\label{eq16}
f_0(v)=\frac{n_p}{\sqrt{2\pi}nv_{th}}e^{-v^2/2v_{th}^2}+\frac{n_b}{\sqrt{2\pi}nv_{T}}e^{-(v-v_b)^2/2v_{T}^2},
\end{equation}
where $n_b$ is the beam density, $n \equiv n_p+n_b$ is the total electron density,  $v_{th}$ and $v_T$ are, respectively, the beam and plasma thermal speeds, and $v_b$ is the beam mean velocity. It is assumed that $v_b \gg v_T$, $v_b \gg v_{th}$ (and in the limit when $v_{th} \rightarrow 0$ and $v_T \rightarrow 0$, $f_0$ writes as a sum of Dirac distributions), and $v_b/v_T \gg (n_b/n_p)^{1/3}$. Moreover, we restrict here to the situation when $n_b \ll n_p$ which strictly corresponds to the beam-plasma instability. Then, it is well-known that the distribution function $f_0$ defined by Eq.~(\ref{eq16}) is unstable, leading to the growth of a nearly monochromatic plasma wave, whose phase velocity is close to $v_b$~\cite{oneil68,oneil71}.  

When the boundary conditions are periodic, the wave amplitude only depends on time. Then, assuming that this wave is sinusoidal, and using the expression Eq.~(\ref{eq2}) for the charge density, Gauss law reads,
\begin{equation}
\label{eq17}
kE_0cos(\varphi)=-(2ne/\varepsilon_0)\left[\langle \cos(\varphi) \rangle \cos(\varphi)+\langle \sin(\varphi) \rangle \sin(\varphi)Ê\right],
\end{equation}
which entails,
\begin{equation}
\label{eq18}
\langle \sin(\varphi) \rangle =0.
\end{equation}

O'Neil, Winfrey and Malmberg addressed the resolution of Eq.~(\ref{eq18}) in Ref.~\cite{oneil71}, by decomposing $\langle \sin(\varphi) \rangle$ as the sum of a term, $\langle \sin(\varphi) \rangle_p$, due to the plasma and a term, $\langle \sin(\varphi) \rangle_b$, due to the beam. Anticipating that the saturation of the instability was due to the beam trapping, they used a linear estimate for $\langle \sin(\varphi) \rangle_p$,
\begin{equation}
\label{eq19}
\langle \sin(\varphi) \rangle_p \approx \frac{\gamma \omega_B^2}{k^3}ÊP.P. \left( \int \frac{f_p(v)-(v-v_\phi)f'_p(v_\phi)} {(v-v_\phi)^3}dv \right),
\end{equation}
where $f_p(v)$ is the first term in Eq.~(\ref{eq16}). This estimate juste amounts to stopping the expansion Eq.~(\ref{eq5}) at first order,  to use the expression Eq.~(\ref{eq6}) for $s_1$  in the limit when $\gamma/k(v-v_\phi) \rightarrow 0$, and to regularize the integral so that it would converge. However, unlike the perturbative expansion detailed in Section~\ref{IIA1}, Eq.~(\ref{eq19}) holds whether the wave grows or decays~\cite{benisti15}. Using Eq.~(\ref{eq19}), the authors of Ref.~\cite{oneil71}  obtained important scaling laws for the beam-plasma instability in the limit when $v_{th}\rightarrow 0$ and $v_T \rightarrow 0$. Nevertheless, in order to obtain actual quantitative results for the nonlinear evolution of the unstable wave amplitude, they had to use particle simulations in order to estimate $\langle \sin(\varphi) \rangle_b$ and solve Eq.~(\ref{eq18}). 

\begin{figure}[!h]
\centerline{\includegraphics[width=12cm]{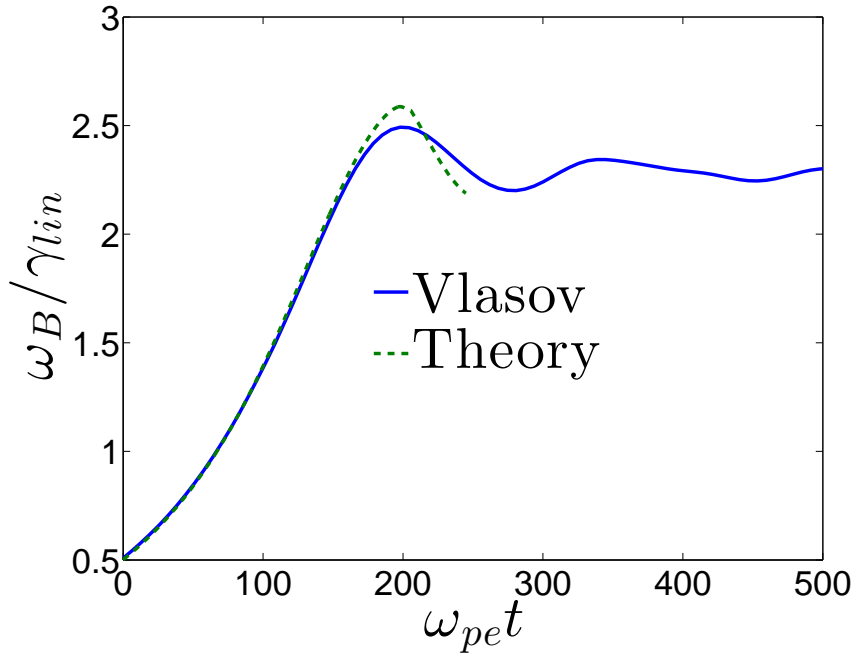}}
\caption{\label{f1} (Color online) Amplitude variation of the unstable plasma wave resulting from the beam-plasma instability as calculated theoretically (green dashed line) and inferred  numerically from Vlasov simulations (blue solid line).}
\end{figure}
Since the publication of Ref.~\cite{oneil71} in 1971, lots of studies have been devoted to the beam-plasma instability. However, we could not find any that would directly solve Eq.~(\ref{eq18}) in a  purely theoretical way. For example,  in Ref.~\cite{carvalero} a statistical approach was used to derive the saturation level of the instability, but the nonlinear growth until saturation could not be addressed. This is precisely what we can do by using the theoretical results of Section~\ref{IIA1}, because the linear growth rate of the instability, $\gamma_{lin}$, is such that $\gamma_{lin}/k(v_b-v_\phi)>1$ [$\gamma_{lin}/k(v_b-v_\phi)=\sqrt{3}$ when $v_{th}=v_T=0$], so that Eq.~(\ref{eq7}) is fulfilled. We theoretically calculate $\langle \sin(\varphi) \rangle$ by deriving its contribution due the plasma from Eq.~(\ref{eq19}), and its contribution due to the beam  from Eq.~(\ref{eq4}) with $f_0$ replaced by the second term in Eq.~(\ref{eq16}). As explained in Section~\ref{IIA1}, before  $S(v_0-v_\phi)$ reaches its first maximum, it is calculated by making use of a perturbative expansion. When the perturbative estimate of $S$ is valid for most electrons in the beam, this lets us express $\langle \sin(\varphi) \rangle_b$ as an explicit function of $\gamma$ and $\omega_B$. Then, Eq.~(\ref{eq18}) becomes an algebraic equation relating $\gamma$ to $\omega_B$, from which we easily derive the nonlinear growth rate of the instability. After $S(v_0-v_\phi)$ has reached its first maximum, its value is derived from Eq.~(\ref{eq14}) and, when this expression applies to nearly all the beam electrons, this yields the explicit time variations of $\langle \sin(\varphi) \rangle_b$. Then, from Eq~(\ref{eq18}) we straightforwardly obtain the time evolution of $\gamma$. Knowing $\gamma$, we derive the nonlinear variations of the wave amplitude, $E_0(t)=E_0(0)\exp\left(\int_0^t \gamma(t')dt'Ê\right)$, and we compare them with those derived from a Vlasov simulation of the beam-plasma instability, using the Vlasov code \textsc{elvis}~\cite{elvis}.  Such a comparison is illustrated in Fig.~\ref{f1} when $n_b/n_p=10^{-3}$, $v_{b}/v_{th}=3.5$ and $v_T/v_{th}=\sqrt{0.02}$. Hence, the beam and the plasma are not infinitely cold and, consequently, the oscillations in the wave amplitude after the first maximum are quickly damped, unlike when $v_T=0$. Moreover, resorting to simulations in order to derive $\langle \sin(\varphi) \rangle_b$ would require calculating the electron motion for a Maxwellian distribution in initial velocities, and not only for a single initial speed as in Ref.~\cite{oneil71}. Theoretically, it is also more complicated to derive $\langle \sin(\varphi) \rangle_b$ when $v_T \neq 0$ than when the beam is infinitely cold. However, in the perturbative regime, $\langle \sin(\varphi) \rangle_b$ is only a function of $\omega/\gamma$ and $kv_{th}/\gamma$ which one may approximate analytically, or at least fit numerically, while using Eq.~(\ref{eq14}) provides values of $\langle \sin(\varphi) \rangle_b$ in a very effective fashion. 

Fig.~\ref{f1} shows that we are able to provide an accurate theoretical prediction for the nonlinear growth and saturation of the beam-plasma instability. Indeed, our estimate for the time at which the first maximum in $\omega_B$ occurs matches that found numerically (they differ by one discretization time), while the theoretical and numerical values of the first maximum differ by about 3.5\%. In Fig.~\ref{f1}, we only plot our theoretical results up to times slightly larger than that corresponding to the first maximum in $\omega_B$.  We do have a theory that reproduces the subsequent oscillations, and that shows that these oscillations are less damped for smaller values of $v_T$. However, such a theory requires additional calculations, which are outside the scope of this paper. A more complete theoretical description of the beam-plasma instability will the subject of a forthcoming article. 
 \subsection{Slowly-varying driven wave in a Maxwellian plasma}
 \label{IIB}
 Let us now consider an EPW growing in an initially Maxwellian plasma, which is only possible if the wave is driven by an external field, whose amplitude is denoted by $E_d$. Then, as shown in Ref.~\cite{vlasovia}, Eq.~(\ref{eq18}) is changed into,
 \begin{equation}
\label{eq20}
-\left \langle \frac{\sin(\varphi)}{\Phi} \right \rangle E_0=E_d,
\end{equation}
where $\Phi\equiv eE_0/kT_e$. 

\begin{figure}[!h]
\centerline{\includegraphics[width=12cm]{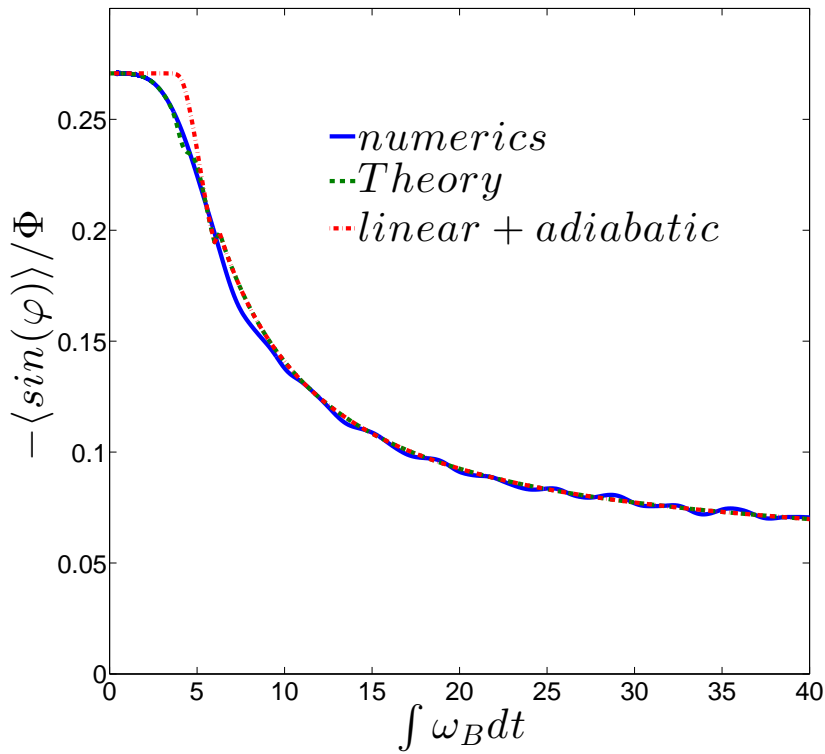}}
\caption{\label{f2} (Color online) Values of $-\langle \sin(\varphi)\rangle/\Phi$ for Maxwellian electrons acted upon by a wave whose amplitude is $E_0(t)=E_0(0)e^{\gamma t}$, with $\gamma/kv_{th}=0.05$, and whose phase velocity is $v_\phi=3v_{th}$. The blue solid line is obtained from test particles simulations, the green dashed line from Eq.~(\ref{eq22}) and the red dashed-dotted line from Eq.~(\ref{eq24}).}
\end{figure}
For small amplitude waves, we still use the perturbative expansion~Eqs.~(\ref{eq4})~and~(\ref{eq5}) to derive $\langle\sin(\varphi) \rangle$. However, when the perturbative estimate of $S(v_0-v_\phi)$ is no longer accurate, we do not make use of the procedure detailed in Section~\ref{IIA1}. Indeed, this would restrict our derivation of $\langle\sin(\varphi) \rangle$ to values of $\omega_B$ of the order of $\gamma$, which would be too small for our theory to apply to the nonlinear modeling of SRS discussed in Section~\ref{IV}. Moreover, the formulas of  Section~\ref{IIA1} provide an accurate expression of each $S(v_0-v_\phi)$, which is not required when dealing with a smooth initial distribution function. Indeed, as discussed in detail in Ref.~\cite{benisti16,fadia}, adiabatic results should be valid whenever $\gamma/k$ is small compared to the typical range in velocity over which $f_0(v_0)$ varies significantly. For a Maxwellian, this reads 
\begin{equation}
\label{ref2}
\gamma/kv_{th}Ê\ll 1.
\end{equation}
However, a direct adiabatic estimate of $\langle \sin(\varphi) \rangle$ would be identically null and, therefore, useless. Heuristically, considering $\gamma$ as the imaginary part of the wave frequency and making a first-order Taylor-like expansion of $\langle e^{-i\varphi} \rangle (\omega+i\gamma)$ yields, 
\begin{equation}
\label{eq21}
\langle \sin(\varphi) \rangle \approx -\gamma \partial_\omega \langle \cos(\varphi) \rangle_a,
\end{equation}
where $\langle \cos(\varphi) \rangle_a$ is the adiabatic estimate of $\langle \cos(\varphi) \rangle$ given by Eq.~(\ref{eq48}) of Section~\ref{IIIB}. Actually, we do not need to resort to a heuristic argument to derive Eq.~(\ref{eq21}). This equation can be proved  by making use of the variational formalism described in Section~\ref{III} and, provided that $\gamma/kv_{th} \alt 0.1$, it is  very accurate whenever $\omega_B/2\gamma \agt 5$~\cite{benisti07}.  As for the perturbative estimate of $\langle \sin(\varphi) \rangle$, by leading the expansion up to a high enough order (we went up to order 11), it remains valid up to values of $\omega_B/2\gamma$ close to 7. Hence, there exists a finite range in $\omega_B/2\gamma$ for which the estimate of $\langle \sin(\varphi) \rangle$ given by Eq.~(\ref{eq21}), and the  perturbative one which we denote by  $\langle \sin(\varphi) \rangle_{per}$, are both accurate. Therefore, one just needs to connect these two estimates about $\omega_B/2\gamma =5$ to obtain a precise expression for $\langle \sin(\varphi) \rangle$, which then reads,
\begin{equation}
\label{eq22}
\langle \sin(\varphi) \rangle = (1-Y)\langle \sin(\varphi) \rangle_{per}+Y\partial_\omega \langle \cos(\varphi) \rangle_a,
\end{equation}
where  $Y\approx 0$ when $\omega_B/2\gamma \ll 5$ and  $Y\approx 1$ when $\omega_B/2\gamma \gg 5$. Actually, $\partial_\omega \langle \cos(\varphi) \rangle_a$ converges very quickly towards $\langle \sin(\varphi) \rangle$ when $\omega_B/2\gamma >5$ so that, in Eq.~(\ref{eq5}), we use for $Y$ a Heaviside-like function. Eq.~(\ref{eq22}) yields a very accurate theoretical expression for $\langle \sin(\varphi) \rangle$ when,
\begin{equation}
\label{eq23}
Y(\omega_B/\gamma)=\left\{\tanh\left[Ê\exp(\omega_B/6\gamma)-1\right]^3 \right\}^8,
\end{equation}
 as may be seen in Fig.~\ref{f2} (and other comparisons made in Refs.~\cite{benisti07,yampo} with other parameters showed the same accuracy). If, instead of using a high order perturbation result for $\langle \sin(\varphi) \rangle_{per}$, we just use a linear estimate, $\langle \sin(\varphi) \rangle_{lin}$, we greatly simplify our expression for $\langle \sin(\varphi) \rangle$, and we still obtain fairly accurate results, as illustrated in Fig.~\ref{f2}  . Hence, for the sake of simplicity, we more conveniently use, 
 \begin{equation}
\label{eq24}
\langle \sin(\varphi) \rangle = (1-Y)\langle \sin(\varphi) \rangle_{lin}+Y\partial_\omega \langle \cos(\varphi) \rangle_a,
\end{equation}
even though the former expression is a bit less accurate than Eq.~(\ref{eq22}). Plugging Eq.~(\ref{eq24}) into Eq.~(\ref{eq20}), one gets the following envelope equation,
\begin{equation}
\label{eq25}
(1-Y)\partial_\omega \chi_{lin}(\partial_tE_0+\nu_LE_0)+Y\partial_\omega \chi_{a}\partial_tE_0=E_d,
\end{equation}
where $\chi_{lin}$ is the linear electron susceptibility, 
  \begin{equation}
\label{eq27}
\chi_{lin}Ê\equiv -\frac{\omega_{pe}^2}{k^2} P.P.\left(\int \frac{f'_0(v)}{v-v_\phi}dv \right),
\end{equation}
and where $\nu_L$ is the Landau damping rate in the limit when it is very small compared to the plasma frequency,
\begin{equation}
\label{eq26}
\nu_L=\frac{-\pi\omega_{pe}^2f'_0(v_\phi)}{k^2\partial_\omega\chi_{lin}}.
\end{equation}
Note that the previous expression for the Landau damping rate is valid whatever the value of $k \lambda_D$ ($\lambda_D \equiv v_{th}/\omega_{pe}$ is the Debye length), because we consider a slowly growing driven wave, and not a freely propagating wave that would decay at the Landau rate. As for the adiabatic susceptibility, $\chi_a$, it is given by Eq.~(\ref{eq49}) of Section~\ref{IIIB}. 

\begin{figure}[!h]
\centerline{\includegraphics[width=15cm]{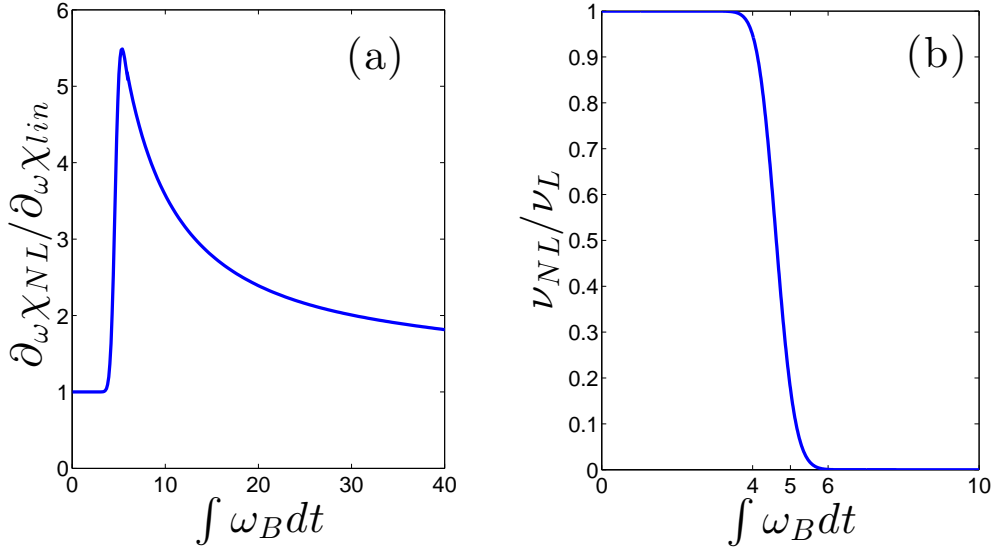}}
\caption{\label{f3} Panel (a), nonlinear variations of $\partial_\omega \chi_{NL}$ and, panel (b), of $\nu_{NL}$.}
\end{figure}
One may also want to write the envelope equation~(\ref{eq25}) the following way,
\begin{equation}
\label{eq28}
\partial_\omega \chi_{NL} \partial_t E_0+\nu_{NL}ÊE_0=E_d,
\end{equation}
where $\partial_\omega \chi_{NL}Ê\equiv (1-Y)\partial_\omega \chi_{lin}+Y\partial_\omega \chi_{a}$ and $\nu_{NL}=(1-Y)\nu_L$. As may be seen in Fig.~\ref{f3}, $\partial_\omega \chi_{NL}$ and $\nu_{NL}$ vary very quickly when $\omega_B/2\gamma \approx 5$ due to the rapid variations in $Y$. This looks like a phase transition for the EPW whose state changes from a linear wave to a wave with trapped electrons. Indeed, for a growing wave such that $d\gamma/dt \ll \gamma^2$, $\omega_B(t)/2\gamma(t) \approx \int_0^t \omega_B(t') dt'$ whenever $\omega_B(t) \gg \omega_B(0)$. Moreover, when $\int_0^t \omega_B(t') dt'>5$, the first trapped electrons have nearly completed one pseudo period of their oscillatory motion in the potential trough, so that trapping is effective. Hence, as regards the nonlinear damping rate,  $\nu_{NL}$, we come to the same conclusion as in the seminal work by O'Neil~\cite{oneil65}, it eventually vanishes due to trapping. However, the variations of $\nu_{NL}$ with $\int \omega_B dt$ are very different from those derived by O'Neil. Instead of oscillations, we find an abrupt decrease. Actually, for a homogeneous wave (or even in a one-dimensional geometry), we do not try to  derive the detailed variations of $\nu_L$. We only come to the conclusion that $\nu_{NL}Ê\approx \nu_L$ when $\int \omega_B dt<4$, and $\nu_{NL}Ê\ll \nu_L$ when $\int \omega_B dt>6$. Hence, we do show the rapid decrease of $\nu_{NL}$ and the phase-like transition of the EPW, but we cannot claim that we have a very accurate theory for the variations of $\nu_{NL}$ when $4 < \int \omega_B dt <6$. These variations, as plotted in Fig.~\ref{f3}, depend on the choice we made for $Y$, which has some arbitrariness. However, this defect in our theory disappears when the wave amplitude varies, at least, in two dimensions. Then, our envelope equation, and in particular the expression we obtain for $\nu_{NL}$, is fairly independent of our choice for $Y$, and we do accurately derive how the nonlinear Landau-like damping rate with the wave amplitude.

\section{Inhomogeneous plasma wave propagation in a non-uniform and non-stationary plasma}
\label{III}
\subsection{Variational formalism and 3-D envelope equation}
\label{IIIA}
The results presented in Section~\ref{IIA2} give the main properties of the electron response to a slowly-varying wave: when $\int \omega_B dt \alt 5$ this response is close to linear and when  $\int \omega_B dt \agt 5$ it may be considered as adiabatic. By a slowly-varying wave, we mean that Eq.~(\ref{ref2}) is still fulfilled where, now, $\gamma$ is the wave growth rate as seen by the electron during their motion. Hence, denoting by $x$ the local direction of propagation of the wave, we only address the situation when, for most electrons, 
\begin{equation}
\label{d2}
(kv_{th})^{-1} [v_\phi \partial_x E_0+v_\bot \partial_\bot E_0]Ê\alt 0.1,
\end{equation}
where $v_\bot$ and $\partial_\bot$ are, respectively, the electron velocity and the derivative along a direction perpendicular to that of the wave propagation. 

Moreover, in a 3-D geometry, $\int \omega_B dt$ is calculated along the electron motion in the wave frame, and its actual value depends on the electron velocity in the direction transverse to that of the EPW propagation. Indeed, in 3-D, the wave is localized within a given space domain, $\mathcal{D}$, that the electrons cross due to their transverse motion. Hence, those electrons with a large transverse speed cross $\mathcal{D}$ in a short time and interact very little with the wave; their response to the EPW is linear. By contrast, the slow electrons experience a slowly-varying amplitude as they cross $\mathcal{D}$, and their response to the wave is adiabatic. Then, generalizing the results of Section~\ref{IIA2}, we conclude that the fraction of adiabatic electrons is,
\begin{equation}
\label{eq29}
Y_{3D}Ê\equiv \int f_\bot(\mathbf{v}_\bot)Y\left( \int \omega_B dt \right)d\mathbf{v}_\bot,
\end{equation}
where $f_\bot$ is the transverse velocity distribution function, normalized to unity. The actual values assumed by $Y_{3D}$ do not depend much on the choice we made for $Y$, except for very small transverse gradients when, typically, $\mathbf{v}_\bot.\mathbf{\nabla}_\bot \omega_B \equiv v_\bot \omega_B/l_\bot \ll d\omega_B/dt$, where the time derivative is calculated along the electron motion. In the opposite limit, when $d\omega_B/dt \approx v_\bot \omega_B/l_\bot$ for most electrons, one may replace $Y(\int \omega_Bdt)$ by $\mathcal{H}(\int \omega_Bdt-5)$ where $\mathcal{H}$ is the Heaviside function, which leads to the following analytic expression for $Y_{3D}$,
\begin{equation}
\label{eq30}
Y_{3D}=1-\exp(-\omega_B^2l_\bot^2/50v_{th}^2).
\end{equation}

Knowing that the fraction $(1-Y_{3D})$ of all electrons respond linearly to the wave, and that the fraction $Y_{3D}$ have a nearly  adiabatic motion, we can generalize the envelope equation~(\ref{eq25}) to a 3-D geometry by resorting to a variational formalism. It is well-known that the Lagrangian density for the electrostatic field and the particles is \cite{benisti16,dodin},
\begin{equation}
\label{31}
L(\mathbf{x},t)=\frac{\varepsilon_0 E_0^2}{2}+\int L_e (\mathbf{x},\mathbf{v},t) f(\mathbf{x},\mathbf{v},t)d\mathbf{v},
\end{equation}
where $f$ is the electron distribution function, and where, for a sinusoidal wave,
\begin{equation}
\label{32}
 L_e (\mathbf{x},\mathbf{v},t) \equiv  \frac{mv^2}{2}-(eE_0/k)\cos(\varphi).
\end{equation}
If we use for $f$ the Klimontovitch distribution function, Lagrange equations are just equivalent to the Newton-Gauss equations. Now, as discussed in Ref.~\cite{benisti16}, if one is only interested in the electrostatic field variations, one may use a Vlasov representation for $f$, which becomes a  function of the field variables only. If $f$ is calculated at zero order in the derivatives of these variables, and of the plasma density, 
 one obtains $L \equiv L(E_0,k,\omega,\varphi,n)$. Then, using Lagrange equations for the field $\varphi$ should yield, for a freely propagating wave, $-\partial_{t\omega}L+\mathbf{\nabla}.\partial_{\mathbf{k}} L=\partial_\varphi L$ i.e., a first order differential equation that allows for the plasma and wave non-stationarity and 3-D non-uniformity. At first sight, this is just the envelope equation we are looking for. However, this equation is not correct because the EPW propagation is dissipative~\cite{note}. Indeed, in the linear regime the wave experiences Landau damping while, as discussed in Section~\ref{IIIAA}, adiabatic trapping entails the irreversible increase of electron kinetic energy, at the expense of the wave. Consequently, the Lagrangian density is nonlocal. As shown in Ref.~\cite{benisti16}, in the linear regime, it depends on the time integral of the electric field while, as discussed in Section~\ref{IIIAA}, in the adiabatic regime the nonlocality of $L$ stems from that of the distribution function. Therefore, one needs to build a nonlocal variational formalism for which there exists no general result. \\

In the linear regime, when $Y_{3D}=0$, the correct envelope equation has been derived in Ref.~\cite{benisti15} and it reads,
\begin{equation}
\label{eq33}
\partial^2_{t\omega}(\chi_{lin}E_0^2/4)-\mathbf{\nabla}.\partial_\mathbf{k}(\chi_{lin}E_0^2/4)+\nu_LE_0^2/2=E_dE_0/2,
\end{equation}
where $\nu_L$ is still given by Eq.~(\ref{eq26}). The envelope equation~(\ref{eq33}) just expresses the variation of the linear plasmon density [see Eq.~(\ref{p4}) of Section~\ref{IVB}]~due to Landau damping and to the external drive. \\

The adiabatic envelope equation, when $Y_{3D}=1$, has been derived in Ref.~\cite{benisti16}~and is,
\begin{equation}
\label{eq34}
-\left[\partial_{t\omega}L_{a}\vert_{\mathcal{A}_s}-\mathbf{\nabla}.\partial_{\mathbf{k}}L_{a}\vert_{\mathcal{A}_s} \right]+\int_0^{v_{tr}} H_t\frac{\mathbf{k}}{k^2}.\mathbf{\nabla} f_\gamma dI=\varepsilon_0 E_0E_d/2.
\end{equation}
In Eq.~(\ref{eq34}), the symbol $\vert_{\mathcal{A}_s}$ means that the derivatives are calculated as though the frozen separatrix was motionless. More precisely, a variation in the wave amplitude or phase velocity may entail a change in the density of the trapped and untrapped electrons, $n_t$ and $n_u$, and the corresponding derivatives in $n_t$ and $n_u$ are \textit{not} to be included in the envelope equation. Consequently, the derivatives inside the bracket in the left-hand side of Eq.~(\ref{eq34}) are not genuine derivatives, and this bracket does~\textit{not}~express a conservation law. Indeed, if there were true derivatives, the bracket in Eq.~(\ref{eq34}) would read $\partial_t \mathcal{A}+\mathbf{\nabla}.(\mathbf{v_g}\mathcal{A})$, where $\mathcal{A}\equiv \partial_\omega L_a$ would be the nonlinear counterpart of the plasmon density, and where $\mathbf{v_g} \equiv -\partial_{\mathbf k}L_a/\partial_\omega L_a$ is the group velocity. Then, the space integral of this bracket would be the time derivative of the nonlinear counterpart of the number of plasmon, $\int \partial_\omega L_a d\mathbf{x}$. Hence, just like in the linear regime, Eq.~(\ref{eq34}) would yield a conservation law (or at least the time variations) for the number of plasmons. By contrast, because the derivatives inside the bracket in the left-hand side of Eq.~(\ref{eq34}) are not genuine derivatives, the space-integral of the left-hand side of Eq.~(\ref{eq34}) cannot be written as an exact time derivative. This equation does~\textit{not}~express a conservation law.

Before entering the detailed definition of each term in Eq.~(\ref{eq34}) we first recall that, as discussed in Ref.~\cite{fadia}, the adiabatic electron distribution function may only be unambiguously defined within each of the following sub-regions of phase space ; region $(\alpha)$ above the frozen separatrix, region $(\beta)$ below the separatrix, and region $(\gamma)$ inside the separatrix. Hence, region $(\gamma)$ contains the trapped electron orbits, so that $I$ in the integral of Eq.~(\ref{eq34}) is nothing but the action defined by Eq.~(\ref{eq10}) of Section~\ref{IIA1}. For untrapped electrons, whose orbits lye in region $(\alpha)$ or $(\beta)$, we use a slightly different definition for the action, 
\begin{equation}
\label{n1000}
I \equiv (2\pi)^{-1} \oint v d\varphi,
\end{equation}
where, again, the integral is calculated over a frozen orbit, corresponding to a given value of the Hamiltonian $H$ defined by Eq.~(\ref{n1}) of Section~\ref{IIA1}. Note that, since $H$ is $2\pi$-periodic, its frozen orbits are closed on the cylinder, and this property is explicitly used in Eq.~(\ref{n1000}) when writing a closed integral over an untrapped orbit. Moreover, in the definition of the action Eq.~(\ref{n1000}), we divide the integral by $2\pi$ (which is the usual convention) while, in Eq.~(\ref{eq10}) for the action of trapped electrons we divided the integral by $4\pi$. This convention was used to avoid a jump by a factor of two in the   definition of the action for orbits just above and just below the separatrix.  For a sinusoidal wave, the analytic expression of $I$ for untrapped orbits is well-known (see Ref.~\cite{benisti07})
\begin{equation}
\label{eq13}
I=\frac{4v_{tr}}{\pi}Ê\sqrt{\zeta}K_2(\zeta^{-1})+\eta v_\phi,
\end{equation}
where $\eta=1$ if the electron orbit is located in region $(\alpha)$, above the frozen separatrix, and $\eta=-1$ if the orbit is in region $(\beta)$, below the separatrix. Moreover, we recall here that $\zeta$ is defined by Eq.~(\ref{eq9}), that $v_{tr}Ê\equiv \omega_B/k$ and that $K_2$ is the elliptic integral of second kind.

Let $f_\alpha$, $f_\beta$ and $f_\gamma$ be, respectively, the electron distribution function in region $(\alpha)$, $(\beta)$ and $(\gamma)$, the adiabatic Lagrangian density is~\cite{benisti16},
\begin{equation}
\label{eq35}
L_a=-\int_{v_{tr}+v_\phi}^{+\infty} f_\alpha(\mathbf{x},I,t) H_udI-\int_{v_{tr}-v_\phi}^{+\infty} f_\beta(\mathbf{x},I,t) H_udI-\int_{0}^{v_{tr}}f_\gamma(\mathbf{x},I,t)H_tdI,
\end{equation}
 with,
  \begin{equation}
\label{eq36}
H_u=\mathcal{E}+mIv_\phi-mv_\phi^2/2,
\end{equation}
and, for a sinusoidal wave,
\begin{equation}
\label{eq37}
\frac{e\mathcal{E}}{kT_e} \equiv \frac{(2-\zeta)\Phi}{\zeta},
\end{equation}
where we recall that $\Phi \equiv eE_0/kT_e$, and where $\zeta$ is related to the action $I$ by Eq.~(\ref{eq13}). In the limit when $\zeta \rightarrow 0$, a simple Taylor expansion would show that $H_u \approx mv^2/2-\Phi_p(x,v,t)$, where $\Phi_p(x,v,t) \equiv e^2 E_0^2/2m(kv-\omega)^2$ is the well-known ponderomotive potential, usually derived by making use of a first-order perturbation analysis (see Ref.~\cite{fst}). Then, $H_u$ is nothing but the fully nonlinear counterpart of the well-known perturbative result.

As for $H_t$, it is defined by,
\begin{equation}
\label{eq38}
H_t=\mathcal{E}-mv_\phi^2/2,
\end{equation}
where, for a sinusoidal wave, $\mathcal{E}$ is still defined by Eq.~(\ref{eq37}) while, now, the relation between $\zeta$ and $I$ is given by Eq.~(\ref{eq11}). 

$f_{\alpha,\beta,\gamma}(\mathbf{x},I,t)$ result from the averaging of the distribution functions in action-angle variables, $\tilde{f}_{\alpha,\beta,\gamma}(\theta,I,t)$, over a $2\pi$-interval in $\theta$ about the considered position $\mathbf{x}$. As discussed in Section~\ref{IIIAA}, these distribution functions are usually nonlocal in the wave amplitude and phase velocity, and they evolve in time in an irreversible fashion. For an initially Maxwellian plasma, this leads to the irreversible increase of the electron kinetic energy, at the expense of the EPW. Consequently, as discussed above, the wave propagation as modeled by Eq.~(\ref{eq34}) is dissipative.  \\

In the general situation, when $0<Y_{3D}<1$, the envelope equation is the weighted sum of Eqs.~(\ref{eq33}) and~(\ref{eq34}),
\begin{eqnarray}
\nonumber
\varepsilon_0 E_0E_d/2&=&\varepsilon_0 (1-Y_{3D})\left\{\partial^2_{t\omega}(\chi_{lin}E_0^2/4)-\mathbf{\nabla}.\partial_\mathbf{k}(\chi_{lin}E_0^2/4)+\nu_LE_0^2/2\right\} \\
\label{eq39}
&&+Y_{3D}Ê\left\{-\left[\partial_{t\omega}L_{a}\vert_{\mathcal{A}_s}-\mathbf{\nabla}.\partial_{\mathbf{k}}L_{a}\vert_{\mathcal{A}_s} \right]+\int_0^{v_{tr}} H_t\frac{\mathbf{k}}{k^2}.\mathbf{\nabla} f_\gamma dI\right\}.
\end{eqnarray}
From Eq.~(\ref{eq39}), the nonlinear collisionless, Landau-like, damping rate of the driven EPW is simply,
\begin{equation}
\label{eq41}
\nu_{NL}=(1-Y_{3D})\nu_L,
\end{equation}
where $Y_{3D}$ is given by Eq.~(\ref{eq29}), and where the Landau damping rate, $\nu_L$, assumes the WKB value~Eq.~(\ref{eq26}). When the analytic formula~Eq.~(\ref{eq30}) for $Y_{3D}$ applies, $\nu_{NL}=\nu_L\exp(-\omega_B^2l_\bot^2/50v_{th}^2)$. Hence, we do provide a precise description of the nonlinear decrease of the EPW damping rate, which is much smoother than in one-dimension (1-D), and which does not exhibit the oscillations derived in the sudden regime addressed by O'Neil~\cite{oneil65}.

\subsection{Nonlocality in the adiabatic distribution function, and dissipation}
\label{IIIAA} 
As discussed in detail in Ref.~\cite{fadia}, even for a slowly varying and nearly periodic dynamics, the action is not conserved due to separatrix crossing. Indeed, from Eqs.~(\ref{eq11})~and~(\ref{eq13}), when an electron whose orbit  lies in region $(\alpha)$ (above the separatrix) gets trapped [i.e., when its orbit moves to region $(\gamma)$, inside the separatrix], its action is shifted by $-v_{\phi_{\alpha\rightarrow \gamma}}$, where  $v_{\phi_{\alpha\rightarrow \gamma}}$ is the wave phase velocity when separatrix crossing occurs. Similarly,  when an electron whose orbit  lies in region $(\beta)$ (below the separatrix) gets trapped, its action varies by $+v_{\phi_{\beta\rightarrow \gamma}}$. Hence,  separatrix crossing entails a change in the electron distribution functions, $f_{\alpha,\beta,\gamma}$, which depend on the whole history of the phase velocity and which are, therefore, nonlocal. These distribution functions can be calculated analytically as a function of the initial one, $f_0$, provided that the typical range in action, $\Delta I$, over which $f_0$ varies is such that $\Delta I >\gamma /k$ (where the growth rate $\gamma$ accounts for the total time variation of the wave amplitude along the electron motion in the wave frame). Indeed, as shown in ~\cite{fadia}, after trapping has occurred, 
\begin{eqnarray}
\nonumber
f_{\gamma}^{>}(I)=	&&f_{\alpha}^{<}(I+v_{\phi_{tr}}) \min\left[\frac{2}{\dot{v}_{\phi_{tr}}/\dot{v}_{tr}+1},1\right]  (1+\dot{v}_{\phi_{tr}}/\dot{v}_{tr}) \\
\label{eq42}
&&+f_{\beta}^{<}(I-v_{\phi_{tr}}) \min\left[\frac{2}{1-\dot{v}_{\phi_{tr}}/\dot{v}_{tr}},1\right]  (1-\dot{v}_{\phi_{tr}}/\dot{v}_{tr}),
\end{eqnarray}
where the superscripts $^{<}$ and $^{>}$ respectively refer to the distribution functions before and after separatrix crossing, and where the upper dot stands for the total time derivative. Moreover, $v_{\phi_{tr}}$ is the wave phase velocity when trapping occurs. Note that an electron initially in region $(\beta)$ with action $I-v_{\phi_{tr}}$  is trapped at the same time and lies on the same trapped orbit as an electron initially in region $(\alpha)$ with action $I+v_{\phi_{tr}}$. On this orbit, the averaged velocity is $\langle v \rangle=v_\phi$, while in region $(\beta)$  $\langle v \rangle<v_\phi$, and in region $(\alpha)$  $\langle v \rangle>v_\phi$. Therefore, if $f_{\beta}^{<}(I-v_{\phi_{tr}})>f_{\alpha}^{<}(I+v_{\phi_{tr}})$, which the case for an initially Maxwellian plasma, trapping has entailed an increase in the electron kinetic energy, at the expense of the wave. 

If the wave amplitude decreases and the electrons get detrapped, the new distribution functions are~\cite{fadia},
\begin{eqnarray}
\label{eq43}
f_{\alpha}^{>}(I)&=&f_{\gamma}^{<}(I-v_{\phi_{dtr}}) \min\left[\frac{1+\dot{v}_{\phi_{dtr}}/\dot{v}_{tr}}{2},1\right]  \frac{1}{1+\dot{v}_{\phi_{dtr}}/\dot{v}_{tr}},\\
\label{eq44}
f_{\beta}^{>}(I)&=&f_{\gamma}^{<}(I+v_{\phi_{dtr}}) \min\left[\frac{1-\dot{v}_{\phi_{dtr}}/\dot{v}_{tr}}{2},1\right]  \frac{1}{1-\dot{v}_{\phi_{dtr}}/\dot{v}_{tr}},
\end{eqnarray}
where $v_{\phi_{dtr}}$ is the wave phase velocity when detrapping occurs. Let us now assume that $\vert \dot{v}_{\phi_{tr}}/\dot{v}_{tr} \vert <1$ and $\vert \dot{v}_{\phi_{dtr}}/\dot{v}_{tr} \vert <1$. Then, plugging Eq.~(\ref{eq42}) into Eqs.~(\ref{eq43}) and (\ref{eq44}) yields, 
\begin{eqnarray}
\label{eq45}
f_{\alpha}^{>}(I+v_{\phi_{tr}})&=&\frac{1}{2}Ê\left\{f_{\alpha}^{<}(I+2v_{\phi_{tr}}-v_{\phi_{dtr}}) \frac{1+\dot{v}_{\phi_{tr}}/\dot{v}_{tr}}{1+\dot{v}_{\phi_{dtr}}/\dot{v}_{tr}}+f_{\beta}^{<}(I-v_{\phi_{dtr}}) \frac{1-\dot{v}_{\phi_{tr}}/\dot{v}_{tr}}{1+\dot{v}_{\phi_{dtr}}/\dot{v}_{tr}}  \right\}\\
\label{eq46}
f_{\beta}^{>}(I-v_{\phi_{tr}})&=&\frac{1}{2}Ê\left\{f_{\alpha}^{<}(I+v_{\phi_{dtr}}) \frac{1+\dot{v}_{\phi_{tr}}/\dot{v}_{tr}}{1-\dot{v}_{\phi_{dtr}}/\dot{v}_{tr}}+f_{\beta}^{<}(I-2v_{\phi_{tr}}+v_{\phi_{dtr}}) \frac{1-\dot{v}_{\phi_{tr}}/\dot{v}_{tr}}{1-\dot{v}_{\phi_{dtr}}/\dot{v}_{tr}}  \right\}
\end{eqnarray}
If, initially, $f_{\beta}^{>}(I-v_{\phi_{tr}})>f_{\alpha}^{>}(I+v_{\phi_{tr}})$, after trapping and detrapping $f_{\alpha}^{>}(I+v_{\phi_{tr}})$ has increased compared to its initial value, and $f_{\beta}^{>}(I-v_{\phi_{tr}})$ has decreased. Consequently, after detrapping the electron kinetic energy is larger than its initial value. Moreover, when $v_{\phi_{tr}} \approx v_{\phi_{dtr}}$, $\vert \dot{v}_{\phi_{tr}}/\dot{v}_{tr}\vert \ll1$, and $\vert \dot{v}_{\phi_{dtr}}/\dot{v}_{tr}\vert \ll1$, $f_{\alpha}^{>}(I+v_{\phi_{tr}})$ and $f_{\beta}^{>}(I-v_{\phi_{tr}})$ are both close to the half sum of the initial distribution functions, so that the electron kinetic energy after detrapping is nearly the same as when the electrons were trapped. Hence, the energy gained by the electrons at the expense of the wave due to trapping is not restored to the EPW after detrapping. Physically this is because, in the near-adiabatic regime, the trapped electrons are very rapidly phase-mixed so that trapped orbits are nearly uniformly populated. On such orbits, one cannot tell whether electrons were trapped from above or from below the separatrix. Consequently, if the wave amplitude decreases while its phase velocity remains nearly constant, it is clear that one half of the electrons would be detrapped above separatrix and one half below (see Refs.~\cite{dissipation,fadia}).  Moreover, from Eqs.~(\ref{eq42})-(\ref{eq46}), it is clear that the change in the electron distribution function due to trapping or detrapping is irreversible. Indeed, if one applies these equations successively, for successive trappings and detrappings, one would never recover the initial distribution functions, even when the wave amplitude decreases back to vanishingly small amplitudes. Hence, the increase of electron kinetic energy due to trapping is irreversible, it entails an irreversible decrease of the electrostatic energy and, therefore, wave dissipation.  Theoretically, this is accounted for in our envelope equation by the fact that the derivatives in the bracket of Eq.~(\ref{eq39}) are not genuine space and time derivatives. Again, the space average of this bracket cannot be written as the exact time derivative of any physical quantity, so that our envelope equation does not express any conservation law. The EPW propagation, as modeled by Eq.~(\ref{eq39}), is dissipative.

In order to describe more precisely how dissipation manifests itself in our theory, let us consider a homogeneous EPW in a uniform and stationary plasma. Then, Eq. (\ref{eq39}) is  equivalent to Eq. (\ref{eq25}) or to Eq. (\ref{eq20}). These equations provide a quantitative estimate of the drive efficiency, which clearly decreases with collisionless dissipation. It is also clear that, for smaller values of $E_d/E_0$, one gets a larger electrostatic field with the same drive amplitude, thus evidencing an increased drive efficiency. Now, there is a sharp contrast between the continuous decrease in $E_d/E_0$ illustrated in Fig.~\ref{f2}, and the abrupt drop in $\nu_{NL}$ plotted in Fig.~\ref{f3} (b). This shows that, once Landau damping has vanished, another means of dissipation is at work, that entailed by adiabatic trapping. 

Moreover, as discussed in Refs.~\cite{vgroup,dissipation}, when the field amplitude depends on space and time, Eq.~(\ref{eq39}) would predict the shrinking of the wave packet during its propagation, even in the limit when $Y_{3D} \approx 1$, so that $\nu_{NL}Ê\approx 0$. This is another example of wave dissipation entailed by trapping.

\subsection{Nonlinear dispersion relation}
\label{IIIB}
The coefficients in the envelope equation~(\ref{eq39}) explicitly depend on the wave frequency and wavenumber. These are derived from the so-called ray equations~\cite{kaufman},
\begin{eqnarray}
\label{r1}
d_t \mathbf{x}_R &=& \partial_\mathbf{k}Ê\Omega \vert_{\mathbf{x,t}}, \\
\label{r2}
d_t\mathbf{K}Ê&=& -\partial_x\Omega \vert_{\mathbf{k,t}},
\end{eqnarray}
where  where $\mathbf{K}(t) \equiv \mathbf{k}[\mathbf{x}_R(t),t]$, and where $\Omega(\mathbf{k},\mathbf{x},t)$ is the function of $\mathbf{k}$, $\mathbf{x}$ and $t$ solving the EPW dispersion relation, so that $\omega(\mathbf{x},t) = \Omega[\mathbf{k}(\mathbf{x},t),\mathbf{x},t]$. Therefore, the envelope equation (\ref{eq39}) needs to be solved together with the EPW dispersion relation. 

The latter equation is derived from Gauss law Eq.~(\ref{eq17}) which leads, at zero-order in the space variations of the field amplitude and for a freely propagating wave,
\begin{equation} 
\label{eq47}
1+\frac{2\langle \cos(\varphi) \rangle}{(k\lambda_D)^2\Phi}=0.
\end{equation}
 In Eq.~(\ref{eq47}), $\langle \cos(\varphi) \rangle$ may be replaced by its adiabatic approximation derived in Refs~\cite{benisti07,dwinho}, 
 \begin{eqnarray}
\nonumber
\langle \cos(\varphi) \rangle_a &\equiv & \int_{v_{tr}+v_\phi}^{+\infty} \left[f_\alpha(I)+f_\beta(I-2v_\phi)\right]\left\{1+2 \zeta \left[\frac{K_2(\zeta^{-1})}{K_1(\zeta^{-1})}-1\right]\right\}dI \\
\label{eq48}
&&+\int_0^{v_{tr}} f_\gamma(I) \left\{-1+2\frac{K_2(\zeta)}{K_1(\zeta)}Ê\right\} dI.
\end{eqnarray}
Indeed, as discussed in Section~\ref{II}, adiabatic results are accurate for large enough amplitudes while, as proved in Ref.~\cite{dwinho}, for a wave growing in a homogeneous plasma,
\begin{equation}
\label{eq49}
\chi_a \equiv \frac{2\langle \cos(\varphi) \rangle_a}{(k\lambda_D)^2\Phi},
\end{equation}
assumes a finite limit very close to the linear susceptibility,  $\chi_{lin}$ defined by Eq.~(\ref{eq26}), when $E_0 \rightarrow 0$. This result remains true if the plasma density is not uniform provided that, as the wave grows, it experiences a density $n(\mathbf{x},t) \equiv \mathcal{N}\left[E_0(\mathbf{x},t)\right]$ such that $d\mathcal{N}/d\sqrt{E_0}$ remains bounded as $E_0 \rightarrow 0$~\cite{dwinho}.  

When the wave is driven, the adiabatic dispersion relation is~\cite{benisti08},
\begin{equation}
\label{eq50}
1+\alpha_d \chi_a=0,
\end{equation}
where $\alpha_d>1$ in the linear limit and rapidly decreases towards unity as the wave amplitude increases. Eq.~(\ref{eq50}) has been solved in Ref.~\cite{benisti08}~for an SRS-driven plasma wave that keeps growing in a uniform plasma, to derive the EPW nonlinear frequency shift, $\delta \omega$. The theoretical values for $\delta \omega$ thus found were compared to results inferred from 1-D Vlasov simulations of SRS, and there was an excellent agreement between numerics and theory. Although the adiabatic distribution functions, $f_\alpha$, $f_\beta$ and $f_\gamma$ are nonlocal, $\delta \omega$ is a local function of the wave amplitude provided that the EPW keeps growing in a uniform plasma. This was numerically evidenced in Ref.~\cite{benisti08} by showing that the amplitude dependence of the wave frequency was the same  at two different positions chosen close to both edges of the simulation box. Indeed, for a wave that keeps growing, $v_{\phi_{tr}}$ in Eqs.~(\ref{eq42})-(\ref{eq46}) is only a function of $E_0$, so that the adiabatic distribution functions may be written as functions of the local wave amplitude. 

Locality is lost if the time variations of the wave are not monotonous, even if the plasma is uniform, because in Eqs.~(\ref{eq42})-(\ref{eq46}) $v_{\phi_{dtr}}\neq v_{\phi_{tr}}$. Then, as shown in Ref.~\cite{dwinho},~there is a hysteresis in the wave frequency, which does not vary the same way with $E_0$ when the amplitude increases as when it decreases. 

Locality is also lost for a wave that keeps growing in a non-uniform plasma, because the adiabatic distribution functions depend on the whole history of the phase velocity, which varies due to nonlinearity and to plasma inhomogeneity. Then, these distribution functions depend on the density profile and on the particular way the EPW grows and propagates in this profile. Only when the variations in the phase velocity entailed by nonlinearity greatly overcome those due to inhomogeneity may local formulas for $\delta \omega$ be relevant~\cite{dwinho}. 

In a multidimensional geometry, due to their transverse motion, the electrons cross the domain $\mathcal{D}$ where the wave is located. Hence, they necessarily experience a non-monotonous wave amplitude, that entails a hysteresis in the wave frequency which needs to be accounted for in addition to the nonlocality due to plasma inhomogeneity. Then, 1-D results as regards the dispersion relation are usually inaccurate~\cite{dwinho}, except if the typical time it takes for the electrons to cross the domain $\mathcal{D}$ is much larger than the time it takes for the wave to saturate.

\section{Application to stimulated Raman scattering in a non-uniform and non-stationary plasma}
\label{IV}

\subsection{Nonlinear kinetic effects on stimulated Raman scattering}
\label{IVA}

Before entering the derivation of the coupled equations modeling SRS, let us recall why one needs to account for kinetic effects. These have been discussed a lot in many papers, especially during the last two decades. In particular, Montgomery~\textit{et al}~measured experimentally Raman reflectivity as a function of the incident intensity of a laser focused in a gas jet, at a position where the density was nearly uniform~\cite{montgomery}. For moderate intensities, they found reflectivities much larger than could be inferred from linear theory (i.e., using Tang formula~\cite{tang}). They dubbed this result ``kinetic inflation'', and it just reflects the fact that an EPW is driven more effectively once collisionless dissipation has been nonlinearly reduced, as discussed in Section~\ref{IIIAA}. From Eq.~(\ref{eq20}) and Fig.~\ref{f2}, it is clear that we are able to quantify very precisely the drive efficiency and, therefore, to address kinetic inflation. Now, reflectivities much larger than those derived  from linear theory have been measured at the National Ignition Facility~\cite{NIC}, so that one cannot ignore kinetic inflation as regards laser fusion. 

\begin{figure}[!h]
\centerline{\includegraphics[width=10cm]{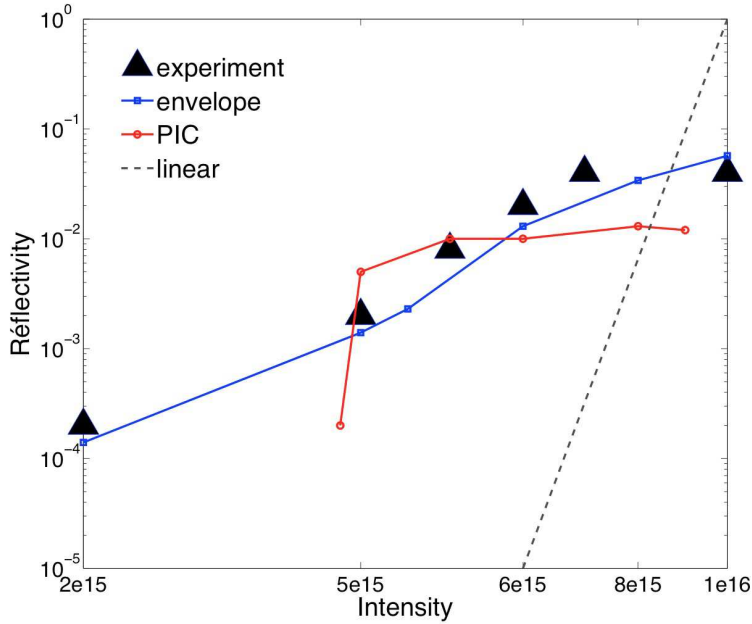}}
\caption{\label{f4} (Color online)  SRS reflectivity as a function of laser intensity. The laser wavelength is $\lambda_l=527$ nm, the plasma density is $n\approx 1.4\times10^{20}$ cm$^{-3}$ ($n/n_c \approx 0.036$), and the electron temperature is $T_e=700$ eV. The space laser profile is assumed to be a Gaussian with a waist $w\approx 2.58$ $\mu$m. The blue squares plot the reflectivities derived from our envelope code~\textsc{brama}~\cite{srs3D,brama}, and the red circles are results from two-dimensional PIC simulations~\cite{yin}. The black dashed line is the reflectivity inferred from Tang formula, and the black triangles reproduce the experimental results of Ref.~\cite{montgomery}. For the experiment, it was estimated that $n\sim1-3 \times10^{20}$ cm$^{-3}$ and $T_e \sim 500$ eV.}
\end{figure}

For larger intensities,  Raman reflectivity was found smaller than inferred from Tang formula, showing that kinetic effects can saturate SRS. Indeed, as discussed in Section~\ref{IIIB}, as the EPW amplitude increases its frequency gets nonlinearly shifted by $\delta \omega$, which may detune the waves coupling~\cite{casanova}. In a multidimensional geometry, $\delta \omega$ assumes a transverse profile that may lead to the EPW self-focusing~\cite{srs3D,yin}. Moreover, a large amplitude EPW is subjected to the trapped-particles instability, leading to the growth of a large band secondary electrostatic field~\cite{friou}. Any of these three effects may lead to SRS saturation. Theoretically, as discussed in Section~\ref{IIIB}, we are able to derive $\delta \omega$ in a very accurate fashion. Moreover, as discussed in Section~\ref{IVC}, we are also able to estimate the impact on SRS of the growth of electrostatic modes resulting from the trapped-particles instability. Therefore, we should be able to address the nonlinear growth and saturation of SRS.

Indeed, as illustrated in Fig.~\ref{f4}, we did manage to reproduce the experimental results reported in Ref.~\cite{montgomery} with our envelope code~\textsc{brama}~\cite{srs3D,brama}. The agreement with the experimental data is at least as good as that obtained with the PIC simulations of Ref.~\cite{yin}, which required computation times larger by a factor of the order of $10^5$ compared to those performed with \textsc{brama}. As described in Refs.~\cite{srs3D,brama}, the code \textsc{brama} solved coupled envelope equations that were only valid when the plasma density was constant, both in space and time. In Section~\ref{IVB}, we generalize these equations so as to account for plasma non-uniformity and non-stationarity, which is one of the main results of this paper.

\subsection{Coupled envelope equations for stimulated Raman scattering}
\label{IVB}
In order to allow for stimulated Raman scattering, the EPW propagation must be solved together with that of an electromagnetic field, which reads, 
\begin{equation}
\mathbf{E}_{em}Ê\equiv \frac{1}{2}Ê\left[-i\mathbf{E}_l e^{i\varphi_l}+\mathbf{E}_se^{i\varphi_s}+c.c.\right],
\end{equation}
where $E_l$ is the electric field amplitude of the incident laser light, and $E_s$ that of the backscattered wave. The laser and scattered frequencies, $\omega_l$ and $\omega_s$, are defined by $\omega_{l,s}Ê\equiv -\partial_t \varphi_{l,s}$, while the laser and scattered wavenumbers are $\mathbf{k}_{l,s}Ê\equiv \mathbf{\nabla} \varphi_{l,s}$. Moreover, $\mathbf{k}_l.\mathbf{k}_s<0$ for a backscattered wave. 

The EPW is coupled with the electromagnetic waves by accounting, in the electron dynamics, for the so-called ponderomotive force, i.e., the longitudinal component (along the waves direction of propagation)  of $-e \mathbf{v}\times \mathbf{B}$, $\mathbf{v}$ being the electron velocity and $\mathbf{B}$ the magnetic field. Then, as discussed in Ref.~\cite{vlasovia}, the electrons are acted upon by an effective electrostatic wave, whose electric field is the sum of that due to the EPW and that due to the ponderomotive force.  Consequently, the electron motion and the charge density may be derived by using the technique discussed in Section~\ref{II}, and Gauss law then leads to Eq.~(\ref{eq39}), where $E_d$ is found to be proportional to $E_lE_s^*$~\cite{vlasovia}. 

The wave equations for the electromagnetic fields directly stem from $c^{-2}\partial_{t^2}Ê\mathbf{A}-\triangle\mathbf{A}=\mu_0 \mathbf{j}_\bot$, where $\mathbf{A}$ is the vector potential and $\mathbf{j}_\bot$ is the transverse current. This latter quantity is approximated by $\mathbf{j}_\bot \approx -(n+\delta n)e^2 \mathbf{A}/m$ (see Ref.~\cite{srs3D}~for details), where $n$ is the unperturbed electron density, and $\delta n = -\varepsilon_0 k E_0/e$. 

Our coupled equations for SRS are written in terms of the following variables. Let us introduce, 
\begin{eqnarray}
\label{p1}
\mathbf{a}_l & \equiv & \frac{\mathbf{E}_l}{\sqrt{\omega_l}},\\
\label{p2}
\mathbf{a}_s & \equiv & \frac{\mathbf{E}_s}{\sqrt{\omega_s}},\\
\label{p3}
\mathbf{a}_p^{lin} &\equiv & \sqrt{\frac{\partial_\omega \chi_{lin}}{2}} \mathbf{E}_0,
\end{eqnarray}
from which we define,
\begin{eqnarray}
\label{p01}
n_l & \equiv & \varepsilon_0\mathbf{a}_l.\mathbf{a}_l/2, \\
\label{p03}
n_s & \equiv &\varepsilon_0 \mathbf{a}_s.\mathbf{a}_s/2, \\
\label{p04}
n_p^{lin} & \equiv & \varepsilon_0\mathbf{a}_p^{lin}.\mathbf{a}_p^{lin}/2.
\end{eqnarray}
$n_l$ and $n_s$ represent the laser and scattered photon density, and $n_p^{lin}$ represents the plasmon density in the linear regime. By analogy with Eq.~(\ref{p04}), we define,
\begin{equation}
\label{p4}
n_p^{NL} \equiv -\partial_\omega L_a. 
\end{equation}
$n_p^{NL}$ cannot be viewed as the nonlinear plasmon density since there is no conservation law for $n_p^{NL}$ in the wave equation~(\ref{eq39}). Indeed, as discussed in Ref.~\cite{benisti16}, there is no way to extend the concept of plasmon to the nonlinear regime when trapping is effective. 

In variables $n_l$, $n_s$, $n_p^{lin}$ and $n_p^{NL}$, our coupled envelope equations for SRS are,
\begin{eqnarray}
\nonumber
\gamma_0 \left\{\mathbf{a}_p^{lin}.\frac{\mathbf{a}_s^*\times\left(\mathbf{k}_l \times \mathbf{a}_l\right)-\mathbf{a}_l \times \left(\mathbf{k}_sÊ\times \mathbf{a}_s^* \right)}{k} \right\}_{tot}&=&[1-Y_{3D}]\left\{\partial_t n_p^{lin}+\mathbf{\nabla}.\left[\mathbf{v}_{gp}^{lin}n_p^{lin}\right] +2\nu_L n_p^{lin} \right\} \\
\nonumber
&&-Y_{3D}\left\{\partial_tn_p^{NL}\vert_{\mathcal{A}_s} +\mathbf{\nabla}.\left[\mathbf{v}_{gp}^{NL}n_p^{NL}\right]\vert_{\mathcal{A}_s} \right\} \\
\label{n7}
&&+Y_{3D} \int_0^{v_{tr}} H_t  \frac{\mathbf{k}}{k^2}.\mathbf{\nabla} f_t dI \\
\label{n8}
-\gamma_0  \left\{a_{p}^{lin}\mathbf{a_s}.\mathbf{a_l} \right\}_{tot} &=&\partial_t n_l+\mathbf{\nabla}.\left[\mathbf{v}_{gl} n_l\right]  -\frac{i\mathbf{a}_l}{2\mu_0\sqrt{\omega_l}}.\triangle_\bot \left(\frac{\mathbf{a}_l}{\sqrt{\omega_l}}\right) \\
\nonumber
\gamma_0  \left\{(a_{p}^{lin})^*\mathbf{a_l}.\mathbf{a_s}  \right\}_{tot}&=&\partial_t n_s+\mathbf{\nabla}.\left[\mathbf{v}_{gs} n_s\right]  -\frac{i\mathbf{a}_s}{2\mu_0\sqrt{\omega_s}}.\triangle_\bot \left(\frac{\mathbf{a}_s}{\sqrt{\omega_s}}\right)\\
\label{n9}
&&+i\left(\delta \omega_c-\frac{\delta \mathbf{k}_c.\mathbf{k}_sc^2}{\omega_s}Ê\right)n_s
\end{eqnarray}
where 
\begin{eqnarray}
\label{p8}
\gamma_0  &=&\frac{\varepsilon_0e k}{m\sqrt{2\omega_l \omega_s \partial_\omega \chi_{lin}}},Ê \\
\label{p9}
\mathbf{v}_{gl}Ê& \equiv & \frac{\mathbf{k}_lc^2}{\omega_l},Ê\\
\label{p10}
\mathbf{v}_{gs}Ê& \equiv & \frac{\mathbf{k}_sc^2}{\omega_s},Ê\\
\label{p11}
\mathbf{v}_{gp}^{lin}Ê& \equiv & \frac{-\partial_{\mathbf{k}}\chi_{lin}}{\partial_\omega \chi_{lin}},Ê\\
\label{p12}
\mathbf{v}_{gp}^{NL}Ê& \equiv & \frac{-\partial_{\mathbf{k}}L_a}{\partial_\omega L_a}.
\end{eqnarray}

Note that, in the envelope equations for SRS, the definition of the EPW electric field has slightly changed. Now, this field is defined by, $\mathbf{E}=-(i/2) \left[\mathbf{E}_0e^{i\varphi}+c.c\right]$, and $\mathbf{E}_0$ is a complex vector. \\

In Eq.~(\ref{n9}), $\delta \omega_c \equiv \omega_l-\omega_s-\omega$ and $\delta \mathbf{k}_c \equiv \mathbf{k}_l-\mathbf{k}_s-\mathbf{k}$. The last term in Eq.~(\ref{n9}) allows for the detuning of the wave coupling entailed by plasma inhomogeneity and by the nonlinear frequency shift. \\

In Eqs.~(\ref{n8}) and (\ref{n9}) we account for wave diffraction, but not in Eq.~(\ref{n7}). Since the laser light is focused at a given point in the plasma, we cannot ignore its diffraction and, similarly, we allow for the diffraction of the scattered wave. We could also account for the EPW diffraction. By analogy with the heuristic procedure leading to Eq.~(\ref{eq21}) in Section~\ref{IIB}, this would require adding the term $(i\varepsilon_0\partial_\omega\chi_{lin}/2)a_p^{lin}\partial_{k_\bot^2}\chi_{lin} \triangle_\bot a_p^{lin}$ to the linear part of Eq.~(\ref{eq39}) [i.e., that part of this equation which is proportional to  Ê($1-Y_{3D}$)]. Similarly, introducing $a_p^{NL} \equiv E_0\sqrt{\partial_\omega \chi_a/2}$, one would need to add the term $(i\varepsilon_0\partial_\omega \chi_a/2)a_p^{NL}\partial_{k_\bot^2}\chi_{a} \triangle_\bot a_p^{NL}$ to the adiabatic part of Eq.~(\ref{eq39}) (i.e., that part of this equation which is proportional to  Ê$Y_{3D}$). However, in the linear regime the phase of the scattered wave imprints on the EPW electric field, so that accounting for the diffraction of the plasma wave does not really  improve the modeling. In the nonlinear regime, $\delta \omega$ induces a phase modulation on the EPW electric field which is much larger than that entailed by the transverse gradient of its amplitude. Again, we conclude that accounting for the diffraction of the plasma wave would not improve the modeling. \\

The envelope equations~(\ref{n7})-(\ref{n9}) need to be solved together with the ray equations. For the plasma wave these are Eqs.~(\ref{r1}) and (\ref{r2}), and for the laser and scattered waves, they are,
\begin{eqnarray}
\label{r3}
d_t\mathbf{x}_{l,s} &=& \mathbf{v}_{g_{l,s}} \\
d_t\mathbf{K}_{l,s}&=&-\nabla \Omega_{l,s},
\end{eqnarray}
 where $\mathbf{K}_{l,s}(t) \equiv \mathbf{k}_{l,s}[\mathbf{x}_{l,s}(t),t]$, and where $ \Omega_{l,s} \equiv \omega_{pe}^2+(k_{l,s}c)^2$.\\
 
 Unlike the fields in the right-hand side of Eqs.~(\ref{n7})-(\ref{n9}), the fields in the left-hand side of these equations are the total fields coming from everywhere in the plasma, and not only those produced locally. This is to account for interspeckle coupling. Indeed, as shown experimentally in Refs.~\cite{glize1,glize2}, the reflectivity of two co-propagating laser pulses is larger than the sum of each reflectivity calculated as though the pulses were propagating alone, because SRS is a collective process. In the experiment of Refs.~\cite{glize1,glize2}, two picosecond pulses propagating in the same direction, but at a transverse distance of about 80 $\mu$m from each other, are focused in a nearly homogeneous plasma at two different times. The intensity of one of the pulses, the so-called strong pulse, is large enough to induce a large reflectivity even when this pulse propagates in the plasma by itself. By contrast, the intensity of the other pulse, henceforth called weak, is too small to lead to any measurable reflectivity when propagating alone. Now, if the weak pulse is launched a few picoseconds (less than about 15 ps) after the strong pulse has interacted with the plasma, it induces a Raman reflectivity of the order of 10\%, i.e., about as much as the strong pulse. 
 
 In Refs.~\cite{glize1,glize2}, three coupling mechanisms have been identified. The  light backscattered from the strong pulse may reach the weak one, and seed SRS (provided that the polarizations of both pulses are the same). Similarly, where the laser and backscattered light of the strong pulse overlap, they produce EPW's that may seed SRS in the weak pulse. Both these effects are accounted for in Eqs.~(\ref{n7})-(\ref{n9}). Now, SRS produces non-Maxwellian electrons which enhance the level of fluctuations or reduce Landau damping as they propagate to nearby speckles. This effect is not accounted for in our envelope equations. However, enhanced fluctuations should be less efficient in triggering SRS than electrostatic or electromagnetic seeding. Indeed, unlike the fluctuations, the waves involved in the seeding just have the right frequency and wavenumber. Moreover, for the reduction of Landau damping to be effective, a majority of electrons should be non-Maxwellian. However, the non-Maxwellian electrons are produced locally in the regions where the laser intensity is largest and, as they escape these regions, they are diluted in the plasma. Therefore we believe that, for a fusion laser smoothed by a random phase plate, where the speckles are close to each other, electromagnetic and electrostatic seeding are the dominant coupling mechanisms.

\subsection{Limits of the modeling}
\label{IVC}
Clearly, the amplitude of an SRS-driven wave is larger where the laser is more intense, in the center of the laser beam (or at the center of a speckle for a smoothed laser). Now, the EPW frequency nonlinearly decreases with its amplitude~\cite{benisti08} and it is, therefore, larger at the edge of the beam than in the center. Consequently, when solving the equations~(\ref{r1}) and (\ref{r2}) one would find that the transverse gradient of the wave frequency would lead to the convergence of the rays i.e., to the EPW self-focusing. When this happens, our hypothesis of a slowly-varying, nearly sinusoidal wave, breaks down. However, since we have an accurate theory for the nonlinear EPW dispersion relation, we are able to predict when self-focusing would lead to SRS saturation. \\

A large amplitude plasma wave is subjected to the trapped-particles instability~\cite{kruer,dodin1} leading to the growth of  secondary electrostatic modes. When the amplitude of these modes is of the same order as that of the EPW,  the hypothesis of a slowly-varying, nearly sinusoidal electrostatic field, breaks down again. Now, from the dispersion relation given in Ref.~\cite{kruer}~or in Ref.~\cite{dodin1}, one can calculate the linear growth rate of the instability,~$\gamma_{SB}$ (which depends on the EPW amplitude, $E_0$), together with the coupled envelope equations for SRS. This allows to estimate the amplitude of the secondary modes, 
\begin{equation}
\label{sb}
E_{SB}(t)=E_{SB}(0) \exp\left\{\int_0^t \gamma_{SB}[E_0(t')]\right\}.
\end{equation}
As shown in Ref.~\cite{friou}, when the amplitude of the fastest growing mode is close to that of the EPW, the plasma wave amplitude stops increasing monotonously, leading to SRS saturation. Hence, we are able to accurately predict the saturation of stimulated Raman scattering entailed by the trapped-particles instability. 

\section{Conclusion}
\label{V}
In this paper, we addressed several issues that pertain to the nonlinear theoretical resolution of the Vlasov-Gauss system (and even of the Vlasov-Maxwell equations when addressing SRS).

The main step of our theory is the derivation of the electron motion under the action of a time-varying electrostatic wave. This derivation stems from the unexpected result that there exists a range in the wave amplitude where a KAM-like perturbation analysis, and neo-adiabatic theory, are both accurate. One just has to connect the results from both  theories, when $\int \omega_B dt \approx 5$, to obtain a theoretical and very accurate description of the electron motion. 

Once the electron motion is known, the charge density is easily calculated and, when it is plugged into Gauss law, it yields the time evolution of the wave. This procedure has been used to provide a purely theoretical derivation of the nonlinear growth and saturation of the beam-plasma instability, and to describe the nonlinear properties of a slowly time-varying EPW in a  Maxwellian plasma with constant density. 

By making use of a nonlocal variational formalism, the latter issue has been generalized to a 3-D geometry and to a non-uniform and non-stationary plasma. The EPW propagation is then modeled by a first order envelope equation. This equation explicitly accounts for the dissipation entailed by trapping, and for Landau damping. In particular, an explicit theoretical expression for the nonlinear Landau damping rate is provided. 

The envelope equation needs to be solved together with the ray equations, which requires solving the EPW dispersion relation. We showed that this dispersion relation could be derived in the linear and nonlinear regimes by resorting to the adiabatic approximation, and that is was nonlocal. Moreover, we gave several examples when this nonlocality could not be ignored.

Finally, by coupling the EPW propagation with those of an incident laser light and of a backscattered wave, we could address stimulated Raman scattering, whose accurate modeling has been a long standing issue in plasma physics. In particular, we derived a set of coupled equations that accounted for nonlinear kinetic effects, interspeckle coupling, and plasma non-uniformity and non-stationarity, and which were valid up to SRS saturation. This is one the main results of this paper.

Solving these equations numerically and interpreting the corresponding results physically is a work in itself, which will be the subject of forthcoming papers.

\begin{acknowledgments}
The author acknowledges David J. Strozzi for providing him with the Vlasov simulation results plotted in Fig.~\ref{f1}.
 \end{acknowledgments}


\begin{thebibliography}{10}
\bibitem{note}
Actually the distinction between the wave and particles is somewhat arbitrary for an electrostatic wave that may only exist due to the spatial arrangement of the particles. Nevertheless, the total particles energy (which is conserved) is of two kinds, kinetic and potential. The large-scale, regular part, of the electrostatic energy is what we call the wave. Now, due to Landau damping, or to trapping, the electron kinetic energy increases in an effectively irreversible fashion. Consequently,  the potential energy, and especially that related to the wave, decreases irreversibly. This is what we call wave dissipation. Now, as regards the wave, its electrostatic energy is not the best quantity to characterize its dissipation. For example,  in the linear regime, Eq.~(\ref{eq33}) of Section~\ref{IIIA} shows the Landau damping of the plasmon density, defined by Eq~(\ref{p4}) of Section~\ref{IVB}. In the nonlinear regime, the situation is more complex because of the difficulty to define the nonlinear counterpart of the plasmon density, as discussed in Ref.~\cite{benisti16}.  Nevertheless, the EPW dissipation is accounted for in our envelope equation, although in a less transparent way than in the simple linear regime.
\end{thebibliography}
\end{document}